\documentclass[12pt,letterpaper]{article}

\usepackage{amsmath}
\usepackage{amsfonts}
\usepackage{amssymb}
\usepackage{graphicx}
\usepackage{epstopdf}
\def\be{\begin{equation}}
\def\ee{\end{equation}}
\def\bea{\begin{eqnarray}}
\def\eea{\end{eqnarray}}

\newcommand{\di}{\partial}

\newcommand{\MP}{M_{\rm Pl}}
\newcommand{\Mbar}{{\,\overline{\vphantom{M}\,\,\,\,\,}\!\!\!\!\!\!M}}

\newcommand{\roughly}[1]%
    {{\mathrel{\raise.3ex\hbox{$#1$\kern-.75em\lower1ex\hbox{$\sim$}}}}}
\newcommand{\lsim}{\mathrel{\roughly<}}
\newcommand{\gsim}{\mathrel{\roughly>}}

\begin{document}

\begin{flushright} {\footnotesize HUTP-06/A0019}\\ {\footnotesize MIT-CTP 3738} \\{\footnotesize IC/2006/034}
 \end{flushright}
\vspace{5mm}
\vspace{0.2cm}
\begin{center}

\def\thefootnote{\fnsymbol{footnote}}

{\Large \bf Starting the Universe:\\[.1cm]
\bf Stable Violation of the Null Energy Condition\\[.1cm]
and Non-standard Cosmologies\\[1cm]}
{\large Paolo Creminelli$^{\rm a}$, Markus A. Luty$^{\rm b}$, Alberto Nicolis$^{\rm c}$, Leonardo Senatore$^{\rm d}$}
\\[0.5cm]

{\small
\textit{$^{\rm a}$ Abdus Salam International Center for Theoretical
Physics\\ Strada Costiera 11, 34014 Trieste, Italy}}

\vspace{.2cm}

{\small
\textit{$^{\rm b}$ Physics Department\\
University of Maryland, College Park MD 20742, USA}}

\vspace{.2cm}

{\small
\textit{$^{\rm c}$ Jefferson Physical Laboratory \\
Harvard University, Cambridge, MA 02138, USA}}

\vspace{.2cm}

{\small
\textit{$^{\rm d}$ Center for Theoretical Physics \\
Massachusetts Institute of Technology, Cambridge, MA 02139, USA
}}

\end{center}

\vspace{.8cm}

\hrule \vspace{0.3cm}
{\small  \noindent \textbf{Abstract} \\[0.3cm]
\noindent
We present a consistent effective theory
that violates the null energy condition (NEC)
without developing any instabilities or other pathological features.
The model is the ghost condensate with the global shift symmetry
softly broken by a potential.
We show that
this system can drive a cosmological expansion with $\dot{H} > 0$.
Demanding the absence of instabilities in this model
requires $\dot{H} \lsim H^2$.
We then construct a general low-energy effective theory that
describes scalar fluctuations about an arbitrary FRW background,
and argue that the qualitative features found in our model are very
general for stable systems that violate the NEC.
Violating the NEC allows dramatically non-standard cosmological histories.
To illustrate this,
we construct an explicit model in which the expansion of our universe originates
from an asymptotically flat state in the past, smoothing out the
big-bang singularity within control of a low-energy effective theory.
This gives an interesting
alternative to standard inflation for solving the horizon problem.
We also construct models in which the present acceleration has $w < -1$;
a periodic ever-expanding universe;
and a model with a smooth ``bounce'' connecting a contracting and
expanding phase.
\vspace{0.5cm}  \hrule

\def\thefootnote{\arabic{footnote}}
\setcounter{footnote}{0}

\newpage
\section{Introduction}
According to our present understanding of cosmology the history
of the universe has been characterized by an incessant slowing
down of the Hubble rate $H = \dot{a}/ a$.
Even the ``accelerated'' expansion ($\ddot{a} > 0$) today
and during inflation at best corresponds to $H = \hbox{\rm constant}$,
and therefore can only halt the decrease of $H$.
Intuitively, the fact that $H$ cannot increase 
is a manifestation of the fact that gravity is generally attractive,
and in some sense never ``too repulsive.''
More formally, $\dot{H} \le 0$ can be viewed as a consequence of the
null energy condition (NEC), which requires that
for all null vectors $n^\mu$ the matter stress-energy tensor
must satisfies $T_{\mu\nu} n^\mu n^\nu \ge 0$.
In a Friedmann-Robertson-Walker (FRW) spacetime this condition reduces
to $\rho + p \ge 0$, which for a spatially flat universe 
directly implies $\dot H \le 0$.

But if the NEC is satisfied and $H$ is always decreasing,
why is the universe expanding in the first place?
The conventional view is that $H$ increases going backward in time
until we reach an era where $H \sim \MP$ (the ``big bang'')
where quantum gravity effects become important.
At this scale general relativity breaks down as an effective theory,
so in this view the mystery of the origin of the expansion of the
universe is inseparable from the issue of the UV completion of gravity.
On the other hand, if the NEC is violated it is possible that
the history of the universe is drastically modified
without quantum gravity effects becoming important.
For example, the present expansion might be due to a smooth ``bounce''
from an earlier contracting phase.
An even more interesting possibility, which is just a limiting case of a bouncing cosmology, is a universe that asymptotically in the past is Minkowski space and as time goes on it simply 
starts expanding connecting smoothly to the subsequent familiar FRW cosmology and thus resolving the Big Bang singularity. 
These possibilities are discussed more fully below.

Another motivation to consider violations of the NEC come from
observations of the current accelerated expansion of the universe.
These are conventionally summarized by the
equation of state parameter $w = p/\rho$.
A cosmological constant has $w = -1$, but present data allow 
$w \gsim -1.2$ \cite{Spergel:2006hy,Seljak:2006bg}.
The NEC implies $w \ge -1$, and it is natural to ask whether
$w < -1$ is a viable possibility.

One simple way to violate the NEC is to add fields with
wrong-sign kinetic terms.
Scalar excitations  have
energy unbounded from below (they are ``ghosts''),
and are therefore subject to catastrophic instabilities at all scales. In particular the vacuum is unstable to decay into ghosts and gravitons
with an infinite rate in any theory that is Lorentz invariant in the UV \cite{riccardo,ghostconstraints2}.
This kind of model can nonetheless be made compatible
with observation by coupling it only to gravity and
postulating that unknown UV physics breaks Lorentz invariance and cuts off the divergence
for short wavelenths \cite{riccardo,ghostconstraints2}.
In this approach, the stability of the universe depends on
unknown UV physics as the leading instability is at
short distances\footnote{See \cite{Holdom:2004yx} for an explicit
  realization of a Lorentz violating cut-off of a ghost instability.}.

The NEC is believed to hold for all well-behaved
systems without instabilities.
Other energy conditions can be violated by
a sensible system simply by adding a suitable
(positive or negative) cosmological constant. The NEC cannot, since it 
is saturated by a cosmological constant.
Indeed it has been shown in Ref.~\cite{fluids} 
for a very broad class of models that whenever the stress-energy
tensor violates the NEC the system has catastrophic instabilities,
either in the form of ghosts or in the form of exponentially
growing modes with arbitrarily short wavelengths (``tachyons'').%
\footnote{The only exceptions are certain anisotropic systems which are
stable but admit superluminal excitations \cite{fluids}.
However such exceptions are not relevant for cosmology, where one is
only interested in isotropic systems.}
Both types of instability occur for all wavelengths down to the UV
cutoff of the theory.
Therefore, unlike the Jeans instability, these instabilities cannot be damped
by Hubble friction, which is only effective at frequencies smaller
than $H$.

The results of Ref.~\cite{fluids} are derived at the two-derivative level
in a systematic low-energy expansion.
Higher derivative corrections are suppressed by some scale $M$,
and in general they can be safely ignored at frequencies and momenta
smaller than $M$.
However this is not the case if the two-derivative action is degenerate,
and some modes propagate only thanks to higher derivative terms.
This is precisely what happens in the ghost condensate
\cite{Arkani-Hamed:2003uy}
and in more general theories of massive gravity \cite{sergio}.
For example, scalar fluctuations around the ghost condensate
have dispersion relation $\omega^2 = \vec{k}^4 / M^2$.
Theories of modification of gravity can be thought as non-trivial
scalar backgrounds with a stress-energy tensor equal to the one of a
cosmological constant \cite{fluids}. Therefore they lie on the threshold of
violating the NEC. This explains the origin of the degenerate
dispersion relations and tell us that these theories are a first step
towards violating the NEC. 

In this paper we use
the ghost condensate as a starting point
to construct a consistent and technically natural low-energy effective field theory
that violates the NEC without any instabilities.
The simplest example is a ghost condensate $\phi$ rolling {\em up} a linear
potential, which gives rise to  superacceleration ($\dot{H} > 0$).
The late-time solution for $\phi$ climbs up the potential indefinitely, making the universe expand with an indefinitely growing Hubble rate, $H \propto \sqrt{t}$.
This solution is an attractor, 
in the sense that trajectories flow to a common late time behavior,
like in conventional
slow-roll inflation.
An interesting feature of this model is that we cannot violate
the NEC by an arbitrary amount, in the sense that we must have
\be
\label{eq:thebound}
\dot{H} \lsim H^2
\ee
to avoid instabilities.
This comes from an interplay between gradient
and Jeans instabilities.
As discussed in Ref.~\cite{fluids}, violating the NEC gives rise to
a dispersion relation of the form $\omega^2 \sim -k^2$, which has
a gradient instability.
This instability is cured at short distances by the
higher-derivative term giving a dispersion relation schematically
of the form $\omega^2 \sim -k^2 + k^4$, which is unstable only
for long times $\omega < \omega_{\rm grad}$.
However, the $k^4$ term increases the mixing with gravity,
leading to a Jeans instability similar to that for ordinary
matter, so the system is unstable for $\omega < \omega_{\rm Jeans}$.
Remarkably, we find the model-independent relation between
the instabilities
\be \label{riccardo}
\omega_{\rm grad} \,  \omega_{\rm Jeans} \sim \dot{H} \;.
\ee
Hubble friction will damp all instabilities only if both
$\omega_{\rm grad}$ and  $\omega_{\rm Jeans}$ are smaller than $H$,
which leads to Eq.~(\ref{eq:thebound}).
%
%
%
We emphasize that
if this inequality is violated, the instability does not extend
to arbitrarily short distances and time scales, and is under
control of the effective theory.
These instabilities are therefore much milder than ghost or tachyon
instabilities, and may in fact have interesting cosmological consequences.
(For example, the conventional Jeans instability gives rise to structure formation
in the universe.)

In fact, these features are much more general than this particular
model.
We demonstrate this by performing a general analysis of scalar fluctuations
about an arbitrary FRW cosmological history.
The analysis can be viewed most simply as the general case of an expansion
driven by a rolling scalar.
However, it also applies to the fluctuations
of the Goldstone mode of time translations in the case where the
expansion is driven by a mixture of fluids and rolling scalar fields,
and so the analysis is quite general.
We find that the qualitative features of the ghost condensate in a linear
potential are completely generic.
In particular, whenever $\dot H$ is positive the system is unstable at the
two-derivative level,
and the same interplay between gradient and Jeans instabilities leads to Eq.~(\ref{riccardo}), so we must have
$\dot{H} \lsim H^2$ in order for instabilities to be completely absent.

We then consider a number of cosmological applications of models based
on ghost condensation that violate the NEC without instabilities.
We first discuss the possibility that the present expansion of the universe
is superaccelerating ($\dot{H} > 0$), corresponding to $w < -1$.
The model consists of the ghost condensate rolling up a potential.
In such a model $H$ can in principle increase smoothly forever,
although at some point $H$ becomes so large that the model exits the
regime of validity of the effective field theory.
To get a measurable violation of the NEC we must be close to saturate the
inequality (\ref{eq:thebound}), suggesting that there may be long wavelength growing modes in the dark energy 
sector that just started evolving. These may provide an additional signal for this class of models.

We then construct a model where the universe starts from
Minkowski space in the asymptotic past.
The model consists of a ghost condensate with $\phi=t$ that rolls up
a potential $V \sim \phi^{-2}$. 
For large negative $t$, this model gives $H \sim |t|^{-1}$,
so the universe goes from a zero curvature, zero energy state to higher curvatures and energies,
and this can then be smoothly connected to a standard FRW
expansion.

Next, we consider a model which gives rise to a cyclic universe
that is always expanding. While the scale factor $a(t)$ steadily grows larger and larger, the Hubble parameter and the energy density are periodic functions of time. 
In this model, a ghost condensate travels along a periodic potential,
giving rise to a phase of superacceleration followed by reheating, a
radiation-dominated and then matter-dominated phase, followed again by superacceleration,
and so on.
In such a scenario, the present accelerated phase is the
beginning of primordial inflation!

Finally, we consider the possibility that the universe smoothly
bounces from a contracting to an expanding phase.
Such a possibility has been considered previously in the literature
(for a review, see Ref.~\cite{justin}),
but in previous treatments the bounce was not under theoretical
control.
We construct a completely smooth bouncing solution with no
instability.
Note that since $H \to 0$ at the bounce, while $\dot{H} \ne 0$,
there is always an unstable mode violating Eq.~(\ref{eq:thebound}).
However, it is easy to arrange that instabilities do not
have time to grow much during the bounce, so that this phase is under theoretical control without catastrophic instabilities.


We conclude that it is possible to have physical systems that
violate the NEC without instabilities or other pathologies,
and that this opens up a large number of interesting new possibilities
for cosmology.
The illustrative examples we consider are intended only as toy models;
we leave the construction of realistic models and their
phenomenology to future work.

\section{\label{example}Superacceleration without Instabilities}
The simplest example of a system that violates the null energy
condition without developing instabilities
can be obtained as a deformation of the ghost condensate
\cite{Arkani-Hamed:2003uy},
as outlined in \cite{Senatore:2004rj}.
Ghost condensation can be realized starting from a Lagrangian for a
derivatively coupled scalar%
\footnote{We use a metric with $(-,+,+,+)$ signature.}
\begin{equation}
\label{eq:Lghost}
{\cal L}= \sqrt{-g}\, M^4 P(X) \;,
\qquad
X = -g^{\mu\nu} \partial_\mu \phi \partial_\nu \phi
\end{equation}
with $M$ an arbitrary mass scale.
(Note that we take $\phi$ to have mass dimension $-1$ so that $X$
is dimensionless.)
The absence of non-derivative couplings is natural if the scalar has a
global shift symmetry
\begin{equation}
\phi \mapsto \phi + \lambda.
\end{equation}
In an expanding universe, one would expect that the field is asymptotically
driven to rest ($\dot\phi\to 0$) by Hubble friction.
However, it is easily checked that there is also a cosmological solution at a minimum of $P$,  
\begin{equation}
\phi=c\, t \; , \qquad P'(c^2)=0 \; ,
\end{equation} 
where the metric is either Minkowski or de Sitter space. This is because at the minima of $P$ the stress energy tensor is the same as for a cosmological constant, although the field is evolving with time \cite{Arkani-Hamed:2003uy}.
Without loss of generality we can take $c=1$ by a redefinition of the field $\phi$. 
We assume that the vacuum energy is positive, so the metric is de Sitter space.
However, unlike a cosmological constant,
the ghost condensate has physical scalar fluctuations defined as
\begin{equation}
\label{NL}
\phi(t, \vec x)=t+\pi(t, \vec x) \;.
\end{equation}
The action for $\pi$ can be obtained expanding the original Lagrangian
Eq.~(\ref{eq:Lghost}).
Up to now we have neglected terms with more than one derivative acting on $\phi$,
like $(\Box\phi)^2$.
Although they are not relevant for the unperturbed solution,
they are important for the $\pi$ dynamics since they give the
leading spatial kinetic term.
We therefore obtain an action of the form 
\begin{equation}
S=\int d^4x\sqrt{-g} \left[
\frac12 M^4\dot{\pi}^2- \frac12 \Mbar^2 \left(\nabla^2\pi\right)^2- \frac12 M^4
\dot{\pi}\left(\nabla\pi\right)^2- \Lambda+\cdots\right] \;,
\label{ghost-action}
\end{equation}
where we have included a cosmological constant term and
chosen $P''(1) = \frac 14$,
which amounts to a redefinition of the mass scale $M$.
The action is not manifestly Lorentz invariant, as expected since the
unperturbed solution spontaneously breaks the Lorentz symmetry.
In particular there is no $(\nabla\pi)^2$ spatial kinetic term,
but the leading term is $(\nabla^2\pi)^2$.
The most generic action for $\pi$ can be obtained using only
symmetry arguments: $\pi$, as shown in Eq.~(\ref{NL}),
non-linearly realizes the broken time diffeomorphisms.
This approach will be developed further below. 

To obtain a model with $\dot{H} > 0$, we introduce a linear potential
\begin{equation}
V = V' \phi,
\qquad
V' = \hbox{\rm constant}\; .
\end{equation}
It is natural to have $|V'| \ll M^5$ because the potential is the only term
that breaks the shift symmetry.
In fact, in the absence of gravity a linear potential does not break the shift
symmetry at all, since ${\cal L} \mapsto {\cal L} + \hbox{\rm constant}$,
and so the only radiative corrections come from gravity.
These are very small given the low cutoff of the effective theory\footnote{Constraints on the cutoff scale,  $M \lesssim 100$ GeV, come from limits on IR modification of gravity  
in ghost condensation \cite{Arkani-Hamed:2003uy,Arkani-Hamed:2005gu}.}, so this form of the effective theory is technically natural.

In the presence of the potential
the background solution changes slightly,
and it can be described by a homogeneous mode $\pi_0(t)$ for the field $\pi$.
Its equation of motion is
\begin{equation}
\ddot{\pi}_0+3H\dot{\pi}_0+ \frac1 {M^4}\frac{d V}{d \phi} =0 \; .
\end{equation}
Neglecting the $\ddot\pi$ term, we find the attractor solution
\begin{equation}
\dot{\pi}_0=-\frac{V'}{3 M^4 H} \label{pizerosol} \;.
\end{equation}
The $\ddot\pi$ term is negligible only if
\begin{equation}
\dot{H} \lsim H^2 \; .
\end{equation}
The velocity of the background is slightly reduced by the tilt as the field rolls up the potential. For small tilt, the stress energy tensor remains close to the one of a cosmological constant with a slowly increasing magnitude, so that $H$ grows with time.  Notice that the $\dot \pi_0$ perturbation becomes smaller and smaller with time as $H$ increases and we approach the following asymptotic behavior
\begin{eqnarray}
&&H^2\simeq \frac{V(\phi)}{3 M_{\rm Pl}^2}  \propto t
\\
&&\dot{\pi}_0 \simeq
 -\frac{V'}{3H(t) M^4}\propto \frac{1}{\sqrt{t}} \;.
\end{eqnarray}
In this solution, $H$ constantly grows with time,
while $\dot\pi_0$ approaches the minimum of the function $P$
(see Fig.~\ref{fig:upthepotential}).


\begin{figure}[th!!]
\begin{center}
\includegraphics[width=14cm]{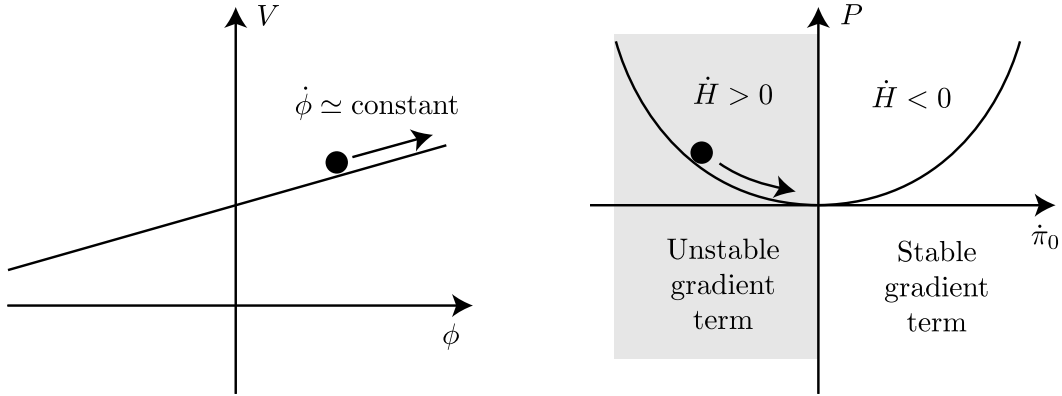}
\caption{\label{fig:upthepotential} \small Schematic
representation of the model presented in section \ref{example}.}
\end{center}
\end{figure}

We now proceed to study the stability of the system.
First of all, we notice that
because we are dealing with an accelerating background
where modes exit the Hubble horizon and freeze out,
an unstable mode has time to grow only if its rate is much faster than $H$.
We can thus restrict to consider wavelengths and timescales much
smaller than $H^{-1}$. 
This means that $H$ and $\dot{H}$ do not change during the time
scales of interest, and can be treated as constants.

The quadratic action for $\pi$ can be obtained expanding 
Eq.~(\ref{ghost-action}) around the new background $\dot\pi_0$: 
\begin{equation}
S=\int d^4x\sqrt{-g}
\left[\frac12  M^4 \dot{\pi}^2+ \dot H M_{\rm Pl}^2
\left(\nabla\pi\right)^2 - \frac12 \Mbar^2 \left(\nabla^2\pi\right)^2 \right]
\; ,
\end{equation}
where we have used that at late times $V'\simeq 6 H \dot{H}M^2_{\rm Pl}$.%
\footnote{There are small shifts of the coefficients of
$\dot\pi^2$ and $(\nabla^2\pi)^2$ that can be safely neglected.}
The background $\dot\pi_0$ induces a
$(\nabla\pi)^2$ term,
and for $\dot H>0$ it has the ``wrong'' sign, corresponding to $P' < 0$.
This signals the presence of instabilities, which we will refer to as `gradient instabilities'. In the absence of a tilt the expansion of the universe drives the field velocity to the point $P'=0$, where there is no $(\nabla\pi)^2$ term. This point separates the stable region $P' > 0$ ---where the fluctuation $\pi$ has a positive gradient energy--- from the unstable one $P'<0$. The negative tilt drives $\dot \phi$ into the unstable regime (see Fig.~\ref{fig:upthepotential}).


Neglecting the mixing of the $\pi$ mode with gravity, we obtain the
dispersion relation
\begin{equation}
\omega^2+ \frac{2 \dot H M^2_{\rm Pl}}{M^4}k^2-\frac{\Mbar^2}{M^4}k^4=0 \;.
\end{equation}
Note that stability is ensured for sufficiently short wavelengths by the
$k^4$ term.
The modes with the fastest instability rate are the ones with
$k^2 \sim  M_{\rm Pl}^2 \dot H/\Mbar^2$ and their rate is of order
\begin{equation} \label{Grad}
\omega^2_{\rm grad}\sim - \left(\frac{\dot{H} M^2_{\rm Pl}}{\Mbar M^2}\right)^2.
\end{equation}
These instabilities are absent if this rate is slower than the Hubble
expansion, which requires the following constraint on the model parameters:
\begin{equation}
\frac{\dot{H}}{H}\lesssim \frac{\Mbar M^2}{M^2_{\rm Pl}}\label{Grad-Constr} \;.
\end{equation}
The mixing of the $\pi$ mode with gravity gives rise to a second kind of
instability,
which is already present in the ghost condensate model in the absence of a potential.
It comes from the mixing of the scalar with gravity and it can be interpreted
as a sort of Jeans instability \cite{Arkani-Hamed:2003uy}.
The instability rate in Minkowski space is of order 
\begin{equation} \label{Jeans}
\omega^2_{\rm Jeans}\sim - \left(\frac{\Mbar M^2}{M^2_{\rm Pl}}\right)^2 \;.
\end{equation}
Also in this case the instability is cured if the Hubble rate is sufficiently fast
\begin{equation}
\frac{\Mbar  M^2}{M^2_{\rm Pl}}\lesssim H \label{Jeans-Constr} \;.
\end{equation}
Comparing Eqs.~(\ref{Grad}) and (\ref{Jeans}), we see that
\be
\omega_{\rm grad} \, \omega_{\rm Jeans} \sim \dot{H} \; ,
\ee
independently of the model parameters.
The two conditions for stability push in opposite directions,
and putting them together
we are left with the window:
\begin{equation}
\frac{\dot{H}}{H}\lesssim \frac{\Mbar M^2}{M^2_{\rm Pl}}\lesssim H \label{condition} \;.
\end{equation}

We conclude that at least for
$\dot{H} \ll H^2$ the model parameters can be chosen in such a way that there are no instabilities,
while $H$ grows indefinitely.
Notice that if Eq.~(\ref{condition}) is satisfied at the initial time,
it remains valid forever as $H$ grows, while $\dot H/H$ decreases with time.




\section{\label{effective}Effective Theory for General FRW Models}
We now turn to a much more general analysis of cosmological
models, and show that the features found in the previous example
are quite general.
In particular, $\dot{H} \lsim H^2$ is always required to avoid
exponentially growing modes in the class of models we consider.

We would like to know whether a given cosmological expansion $a(t)$
suffers from instabilities.
Clearly, we need to know all the light degrees of freedom to answer
this question, and we cannot answer it in complete generality.
We will consider a framework which we believe is very general.
Most simply, it can be described as the effective theory for fluctuations about a FRW background driven  by a single rolling scalar
with a completely arbitrary Lagrangian.
This therefore includes for example a standard slow-rolling inflaton and all possible deformations of the ghost
condensate, such as the one considered in the previous section.

%


In fact, the analysis is even more general.
We are effectively focusing attention on
a scalar excitation which is present in virtually all expanding universes
independently of what matter is actually driving the expansion: the perturbation corresponding to
a common, local shift in time for all the matter fields $\psi_m$.
That is, given a background homogeneous FRW solution
$a(t)$, $\psi_m (t)$, we consider the perturbation
\be \label{perturbation}
\delta \psi_m (x) \equiv \psi_m(t+\pi(x)) - \psi_m(t) \; ,
\ee
parametrized by $\pi(x)$, and the corresponding scalar perturbation of the metric as imposed by Einstein's equations.%
\footnote{Of course one such perturbation is trivial, corresponding
to the FRW background solution with unperturbed matter fields expressed
in an unconventional set of coordinates.
We eliminate this pure gauge mode by fixing a gauge for the metric.}
This is the perturbation which in the long-wavelength limit is called
{\em adiabatic} and obeys a conservation law which is insensitive to
the matter content of the universe \cite{weinberg}.
By its very definition such a perturbation can be gauged away from
the matter sector by an $x$-dependent shift in time.
In other words, as long as we are only interested in this specific perturbation
we can always choose a gauge in which the matter fields are unperturbed,
$\delta \psi_m = 0$,  and the scalar fluctuation is in the metric.
We will adopt this gauge choice and refer to it as `unitary gauge'.
(For the case where the expansion is driven by a single scalar
field, this corresponds to the gauge choice $\phi(t, \vec{x}) = \phi_0(t)$,
where $\phi_0(t)$ is the unperturbed scalar solution.)

The existence of such a fluctuation mode sounds like a trivial consequence
of the presence of a time-dependent FRW background, but it is not.
If the expansion of the universe is driven by a mixture
of rolling scalar fields, then the fluctuation we are interested in is indeed
defined by Eq.~(\ref{perturbation});
it is the Goldstone boson associated to the spontaneous breakdown
of time-translations.
However in general the situation is more subtle.
For instance the ground state of
a solid or a fluid is characterized by three scalar condensates
$\langle \phi^i (x) \rangle = x^i$ ($i=1,2,3$) that
spontaneously break the product of spatial translations and internal shift symmetries
down to the diagonal subgroup \cite{fluids}; in particular time translations are unbroken.
Still, in the case of a fluid it can be shown that non-vorticous excitations---i.e., sound waves---can be described at the classical level as Goldstone bosons of broken {\em time}-translational invariance around a rolling scalar
condensate $\langle \phi(x) \rangle = t$, like a ghost-condensate \cite{fluids}.
So for any mixture of cosmic fluids and rolling scalars there exists an excitation that behaves like the Goldstone of broken time-translations, whose action is invariant under spatial translations and rotations. When
coupling to gravity is taken into account, this translates into invariance under time-dependent spatial diffeomorphisms
$x^i \to x^i + \xi^i(\vec x,t)$.
For more generic systems,
e.g.~solids or Lorentz-breaking massive gravity models \cite{sergio},
the low-energy degrees of freedom and the residual
symmetries will be different, and our conclusions will not necessarily apply. Notice moreover that the presence of additional modes besides the one we are considering could, by mixing with 
it, change its dynamics.

We now proceed to construct the most general effective action for the Goldstone, around a generic FRW background $a(t)$. As we discussed, in unitary gauge the scalar mode does not appear
explicitly in the action but it is part of the metric.
Once we choose unitary gauge in fact time reparameterizations are not allowed anymore so that the metric contains an additional scalar degree of freedom. As we will discuss later the
$\pi$ dependence of the action can be restored using the usual St\"u{}ckelberg procedure, as in the case of massive gravity \cite{Arkani-Hamed:2002sp}.  In unitary gauge the full 
dynamics is described by an action for 
gravity which does not have the full diffeomorphism invariance, but it is only invariant under time-dependent spatial diffeomorphisms $x^i \to x^i + \xi^i(\vec x,t)$. 

It is particularly convenient to work with ADM variables \cite{Arnowitt:1962hi}.
These are the `lapse'
$N \equiv 1/\sqrt{-g^{00}}$, the `shift' $N_i \equiv g_{0i}$,
and the induced metric $\hat g_{ij}$ on hypersurfaces of constant $t$.
In the following we will lower and raise spatial indices with the
three-dimensional metric $\hat g _{ij}$ and its inverse $\hat g^{ij}$.
In ADM variables the full 4D metric reads
\be
ds^2= -N^2 dt^2 + \hat g_{ij} (dx^i + N^i dt) (dx^j + N^j dt) \;.
\ee
The ADM formalism keeps manifest the invariance under 3D space diffeomorphisms: only quantities which are manifestly covariant under
these transformations appear in the equations. The invariance under time reparameterization, although not manifest, is obviously still there. 
For our system, in unitary gauge the time variable is set by the unperturbed matter fields $\psi_m(t)$ so that the splitting between time and space, which is in general arbitrary,
takes here a physical meaning. Therefore the unitary gauge action is {\em not} invariant
under time diffeomorphisms, while all the unbroken symmetries (time-dependent spatial diffeomorphisms) are manifest in the ADM language.

The Einstein-Hilbert action is expressed in the ADM language as 
\be \label{EH}
S_{\rm EH} = \frac 12 M_{\rm Pl}^2 \int \! d^4 x \: \sqrt{-g} \, R = \frac 12 M_{\rm Pl}^2 \int \! d^3 x \, dt  \: \sqrt{\hat g} \, \big[ N  R^{(3)}
+ \frac{1}{N} (E^{ij} E_{ij} - E^i{}_i {}^2) \big] \; ,
\ee
where $R^{(3)}$ is the Ricci scalar of the induced 3D metric. $E_{ij}$ is related to the extrinsic curvature $K_{ij}$
of hypersurfaces of constant $t$,
\be \label{extrinsic}
E_{ij} \equiv N K_{ij} = \frac 12 [{\partial_t {\hat g}}_{ij} - \hat \nabla_i N_j - \hat \nabla_j N_i]
\; ,
\ee
where $\hat \nabla$ is the covariant derivative associated to the
induced 3D metric $\hat g_{ij}$.

The full action consists of the Einstein action plus matter terms.
The Einstein action is invariant under time reparameterizations,
but the matter action is not because we have chosen the gauge
$\pi \equiv 0$ where the scalar fluctuations are parameterized by
the metric.
We expand about a given cosmological background
\begin{equation}
ds^2 = -dt^2 + a^2(t) d \vec{x}^2 \; ,
\quad
\psi_m = \psi_{m 0}(t)
\end{equation}
and consider general metric fluctuations
\begin{equation}
N = 1 + \delta N \; ,
\quad
N_{j} = \delta N_{j} \; ,
\quad
\hat{g}_{ij} = a^2(t) \delta_{ij} + \delta \hat{g}_{ij} \; .
\end{equation}
The action written in terms of the fluctuation fields will have
time-dependent coefficients because they are functions of the
background matter fields $\psi_{m 0}(t)$.
We therefore take the matter action to consist of the most
general Lagrangian invariant under spatial diffeomorphisms,
with time-dependent coefficients.
%
%
Although the natural integration measure is
$\sqrt{-\hat g}\, d^3 x \,dt $, we find it more convenient to use the
4D invariant measure $\sqrt{-g} \, d^4 x = \sqrt{\hat g}\,  N \, d^3x\,  dt$.
Since $N$ is a scalar under spatial diffeomorphisms, this simply amounts to a
reshuffling of the terms in the Lagrangian.

\subsection{Two-Derivative Goldstone Action}
In this subsection, we use the formalism described above to construct
the two-derivative action for the $\pi$ mode described in the previous
subsection.
At short distances and times where the mixing with gravity can be neglected,
this describes the physical scalar fluctuation of the system.
The results obtained here will be confirmed by a complete analysis of the
full gravitational action in the section \ref{Paolo}, but the present analysis
is much simpler and more transparent.

Our strategy is to write the most general matter action in unitary gauge
$\pi \equiv 0$
order by order in the metric fluctuations $\delta N$, $\delta N_{j}$, and
$\delta\hat{g}_{ij}$.
We can then restore the $\pi$ dependence using the St\"u{}ckelberg trick.
That is, given a space-diff invariant term in unitary gauge, e.g.~${1}/{N^2}$, we write its transformation law under {\em time}-diffeomorphisms
\be \label{timediff}
\frac{1}{N^2} = g^{00} \mapsto \frac{1}{N^2}
+ 2 \, \partial_0 \xi^0 - (\partial_\mu \xi^0)^2 \; ,
\ee
where we kept terms of zeroth order in the metric fluctuations and of second order in the transformation parameter $\xi^0$.
We can then restore invariance under time reparameterizations by introducing
the field $\pi$ transforming as
\begin{equation}
\pi \mapsto \pi - \xi^0 \; ,
\end{equation}
and using the gauge-invariant combination ${1}/{N^2}+ 2 \, \partial_0 \pi - (\partial_\mu \pi)^2$ in place of ${1}/{N^2}$ wherever the latter appears in the unitary-gauge Lagrangian. This procedure, where we neglected the gravitational perturbations, gives the action for the Goldstone
$\pi$.
In order to expand the action at quadratic order in $\pi$
we need the transformation law of all terms in the action under
time diffeomorphisms at second order in the transformation
parameter $\xi^0$. The Einstein-Hilbert action is invariant under
a generic 4D diffeomorphism, thus from there we get no
contribution to the $\pi$ Lagrangian. The same holds true for the
measure $\sqrt{-g} \, d^4 x$ in the matter action.
Also all coefficients that explicitly depend on time
should be evaluated at $(t+\pi)$. However it is reasonable to assume that the typical time scale for these coefficients will be of order $H^{-1}$ so that, 
after expanding in $\pi$, these give rise only to non-derivative terms suppressed by $H$, $\dot H$, etc.: we can thus neglect these contributions 
as long as we are interested in
fluctuations with wavelengths much shorter than the cosmological
horizon.
Finally, we are interested in constructing the action for $\pi$ in a systematic derivative expansion. In unitary gauge this corresponds to an expansion in derivatives of the metric fluctuations, which we now proceed to write down.

In order for the given FRW background to solve Einstein's equations, the matter action must contain `tadpole' terms, i.e.~terms that start linear in the metric fluctuations.
Note that the linear terms are canceled by a tadpole term
in the Einstein action, since we are expanding about a solution of the
full action.
However, the Einstein action is completely invariant under time
reparameterizations while the matter action is not, so it is useful to keep the linear terms in the matter action for the St\"u{}ckelberg trick.
At the zero-derivative, linear level the only
invariant under spatial diffs is $\delta N$. Apart from that, a linear piece in the fluctuation can also come from the $\sqrt{-g}$
that makes the integration measure diff invariant. Therefore we have two independent operators at this order.
Instead of choosing the operators $\delta N$ and 1 as our `basis', we find it more convenient to choose a different combination, $1/N^2$ and $1$; the two bases are equivalent at linear order, their difference being quadratic in $\delta N$; the reason of this choice will soon become clear. At this order the matter action is thus of the form
\be \label{tadpoles}
S_{\rm matter} = \int  \! d^4 x  \: \sqrt{- g}
\left[ c(t) \frac{1}{N^2} -  \Lambda(t) \right] \; .
\ee
As discussed above, we allow for generic functions of time $c(t)$,
$\Lambda(t)$ as coefficients of the operators in the Lagrangian.
In fact, the above action, truncated exactly at this level,
is the complete action in the case of an ordinary scalar field $\phi$ with a
potential $V(\phi)$ in unitary gauge.
To see this, just re-write the standard action for $\phi$ on a solution
$\phi (t)$ using ADM variables:
\be
S_{\phi} = \int \! d^4x \: \sqrt{-g}
\left[ -\frac 1 2 (\partial \phi)^2 - V(\phi) \right] =
\int \! d^4x \: \sqrt{- g} \left[ \frac 1 {2 N^2} \dot \phi^2 - V(\phi(t)) \right]  \; ,
\ee
which is exactly of the above form with $c(t) = \frac12 \dot \phi^2$ and
$\Lambda(t) = V(\phi(t))$.
This is the reason why the parameterization in Eq.~(\ref{tadpoles}) is convenient.

In principle one could also add tadpole terms that involve derivatives of the metric,
like for instance
an extrinsic curvature term $K^i {}_i$. However by integration by parts such terms can always be rewritten
as a combination of our non-derivative tadpoles plus derivative terms that involve higher powers of
metric fluctuations, which we will discuss later.

We now rewrite the coefficients $c(t)$ and $\Lambda(t)$ in terms of the Hubble
parameter and its derivatives using the Friedmann equations.
The matter stress-energy tensor evaluated on the background configuration can
be read from the action Eq.~(\ref{tadpoles}):
\be
T_{\mu \nu} = -\frac{2}{\sqrt{-g}}\frac{\delta S_{\rm matter}}{\delta g^{\mu\nu}}
\quad \Rightarrow \quad \left\{ \begin{array}{l}
T_{00} = c(t)+\Lambda(t) \\
T_{ij} = a^2(t) \delta_{ij} \big( c(t) - \Lambda(t)\big)
\end{array} \right.
\ee
Therefore Friedmann equations are
\begin{eqnarray}
H^2 & = & \frac{1}{3 M_{\rm Pl}^2} \big[ c(t)+\Lambda(t)\big]  \\
\frac{\ddot a}{a} & =  & -\frac{1}{3 M_{\rm Pl}^2} \big[ 2 c(t)-\Lambda(t) \big]
\end{eqnarray}
Note that quadratic and higher order terms in the matter action do not contribute to the background Einstein's equation, since their contribution to the stress-energy tensor is at best linear in the fluctuations and therefore vanishes if computed on the background.
Thus the above Friedmann equations uniquely determine the coefficients $c(t)$, $\Lambda(t)$. Eq.~(\ref{tadpoles}) can therefore be written
\be \label{tadpoles_2}
S_{\rm matter} = \int  \! d^4 x \: \sqrt{-g} \left[
- M_{\rm Pl}^2 \dot H \frac{1}{N^2} - \MP^2
(3 H^2 + \dot H) + \cdots \right] \; .
\ee
We now make use of the St\"uckelberg trick. Given the transformation law Eq.~(\ref{timediff}), we obtain simply
\be \label{S_pi_tadpole} S_{\rm matter} \to
S_\pi = \int  \! d^4 x \: \sqrt{-g} \, (M_{\rm Pl}^2 \dot H) (\di
\pi)^2 \; , 
\ee
where we neglected the total derivative linear
term $\di_0 \pi$.%
\footnote{This term is a total derivative as long
as we consider momenta much larger than $H$. When the time-dependence of the coefficients is completely
taken into account all linear terms are going to
cancel exactly, since the background solves Einstein's equations.}

The Lagrangian for $\pi$ is that of a standard relativistic
scalar, but with an overall $-\dot H$ factor.
Note that whenever $\dot H > 0$ the kinetic term has the wrong sign,
leading to a catastrophic ghost instability.


We can try to cure the instability we just found by adding quadratic terms to the effective action Eq.~(\ref{tadpoles_2}). At the zero-derivative level in ADM variables the only possible quadratic  operator is $(\delta N)^2$,
\be
S_{\rm matter} \to S_{\rm matter} + \int  \! d^4 x  \, \sqrt{ -g} \,
\frac12 M^4(t)\, (\delta N)^2 \; .
\ee
The coefficient $M^4(t)$ is unconstrained, and it can be a generic function of time. In realistic situations we expect its typical
time-variation rate to be of order $H$, so we can approximate it as a constant when studying short-wavelength fluctuations.
In order to reintroduce the Goldstone field we just need the trasformation law of $N$ at first order in $\xi^0$,  $N \mapsto N-\di_0 \xi^0$.
This changes the Goldstone action by
\be \label{pi_dot}
 S_\pi \to S_\pi + \int  \! d^4 x \, \sqrt{-g} \; \frac12 M^4 \, \dot \pi ^2 \; .
\ee
As already stressed in Ref.~\cite{Hsu:2004vr}, for large enough $M^4$---larger than $ M_{\rm Pl}^2 \dot H$---the $\dot \pi^2$ gets the healthy sign,
but the gradient part of the action still has the wrong sign for $\dot H > 0$,
signaling the presence of exponentially growing gradient
instabilities at all wavelengths.

At the zero-derivative level in the metric, which corresponds to the two-derivative level in the Goldstone, there is nothing more we can do. Indeed in Ref.~\cite{fluids} it was shown in a broad class of theories that whenever the matter stress-energy tensor violates the null energy condition the system has instabilities, either in the form of ghosts or in the form of exponentially growing modes. This was shown at the two derivative level. The example of the last section makes no exception, its stability crucially relying on higher derivative terms in the Goldstone Lagrangian and on the presence of the horizon for cutting off the instability in the IR. We are therefore led to consider higher-derivative terms in the effective action.

\subsection{Higher-derivative Goldstone Action}
%
%
At next order in the derivative expansion we must include terms that involve the extrinsic curvature $K_{ij}$ of constant-$t$ hypersurfaces. Of course, since $N$ is invariant under spatial diffs, we can as well use its `reduced' version $E_{ij}$ of Eq.~(\ref{extrinsic}).
Notice that the background solution has non-zero extrinsic curvature, $E_{ij}^{(0)} = a^2 H \, \hat g_{ij}$, so we are interested in the fluctuation $\delta E_{ij} \equiv E_{ij}-E_{ij}^{(0)}$. With it we can construct the quadratic operators $\delta E^i {}_i \, \delta N$, $\delta E^{ij} \, \delta E_{ij}$, and $\delta E^i {}_i {}^2$.
Let us ignore for the moment the first of these operators, and let us concentrate on the other two, for which
the discussion is simpler.
In order to see what these operators correspond to in terms of the $\pi$ field, it is sufficient to consider the transformation law of $\delta E_{ij}$ at linear order in the time-diff parameter $\xi^0$ and at zeroth order in the metric fluctuation. From the trasformation law $g_{\mu\nu} \to g_{\mu\nu} + \nabla_\mu \xi_\nu +\nabla_\nu \xi_\mu$ we get
\be
\delta E_{ij} \to \delta E_{ij} - \delta_{ij} \,  \di_t (a^2 H  \xi_0) - \, \di_i \di_j \xi_0 \; .
\ee
The term proportional to $\delta_{ij}$ is negligible for momenta much larger than $H$.
In such a limit we have
\be \label{di_pi}
\int  \! d^4 x \: \sqrt{- g} \,\left[ -\frac12 \tilde M^2 \, \delta E^i {}_i {}^2 - \frac12
\tilde M'^2 \, \delta E^{ij} \delta E_{ij} \right]
\to
\int  \! d^4x \: \sqrt{-g} \left[ - \frac12 \Mbar ^2 \, \frac{1}{a^4}(\di_i^2 \pi)^2
   \right] \; ,
\ee
where $\Mbar ^2 \equiv \tilde M^2 +\tilde M'^2 $, and as before we approximated the
coefficients $ \tilde M^2$ and $ \tilde M'^2$ as constant in time.

According to our effective field theory approach, we are assuming that the scales $M$, $\tilde M$,
and $\tilde M'$ are all of the same order of magnitude, and all further terms appearing in the effective
action will be weighted by the same scale. Then terms that involve higher time-derivatives on $\pi$
will be negligible with respect to Eq.~(\ref{pi_dot}) at frequencies below this cutoff scale $M$.
Similarly, terms involving more spatial derivatives on $\pi$ can be neglected with respect to
Eq.~(\ref{di_pi}) at momenta smaller than $M$.
Therefore in this regime the only possible other term in unitary gauge is a linear term in the 3D curvature $R^{(3)}$.
However, as clear from Eq.~(\ref{EH}) such a term can always be re-written as an Einstein-Hilbert term plus
terms quadratic in $E_{ij}$.

Collecting all of the terms considered so far, the matter action is
\be 
\label{theeffectiveaction}
S_\pi = \int \! d^4 x \:
\sqrt{-g}\left[ \left(\frac{M^4}{2}-M_{\rm Pl}^2 \dot H \right)
 \, \dot \pi^2 + (M_{\rm Pl}^2
\dot H) \, \frac 1{a^2} (\di_i \pi)^2 - \frac{\Mbar ^2}{2} \,
\frac{1}{a^4}(\di_i ^2 \pi)^2 
\right] \; .
 \ee
The
positivity of the $\dot \pi^2$ term is guaranteed for
$M^4- 2 M_{\rm Pl}^2 \dot H >0$.
For simplicity in the rest of this section we assume $M^4-2 M_{\rm Pl}^2
\dot H \simeq M^4$.
Then the dispersion relation of $\pi$
excitations is exactly the one we studied in the example of section~\ref{example}
 \be 
 \label{dispersion} 
 \omega^2 = - \frac{2 M_{\rm Pl}^2 \dot H}{M^4} k^2 + \frac{\Mbar ^2}{M^4}  k^4 \; , 
\ee 
where $k$ is the physical momentum.
For positive $\dot H$, the system has oscillatory
excitations with dipersion law $\omega^2 \propto k^4$ at large
momenta and exponentially growing instabilites at low momenta,
$\omega^2 \propto -k^2$. That is, the higher derivative term cures
the instability at small wavelengths. The critical momentum below
which  the system is unstable is \be k_{\rm grad} \sim   \dot H^{1/2}\frac
{M_{\rm Pl}}{\Mbar} \; , \ee corresponding to a typical instability
rate \be \omega_{\rm grad} \sim \dot H \frac {M_{\rm Pl}^2}{\Mbar M^2}\; . \ee 
We are interested in pushing the
instability time-scale outside the cosmological horizon, where the
Hubble friction freezes the dynamics of perturbations. Note that
the critical length-scale can even be much smaller than the Hubble
horizon---what matters is the typical time-scale for developing
the instability.

Up to now there is no obstacle in making the instability rate as
slow as we like. It is enough to choose a very large $\Mbar^2$
coefficient for the higher derivative term to make $\omega_{\rm grad}$ much
smaller than $H$. But here is where the mixing with gravity
becomes crucial. Usually, if we restrict to momenta and
frequencies much larger than $H$ we are allowed to neglect the
fact that matter fluctuations are mixed with gravitational ones.
This is because, by definition of $T_{\mu\nu}$, such a mixing
comes from a Lagrangian term of the form $h_{\mu\nu} \, \delta
T^{\mu\nu}$, where $\delta T^{\mu\nu}$ is the fluctuation in the
matter stress-energy tensor at first order in the matter
fluctuations. Usually this is non-zero only if the {\em
background} stress-energy tensor is non-zero. For instance, for an
ordinary massless scalar $\phi$ one has \be h_{\mu\nu} \delta
T^{\mu\nu} \sim h \, \dot\phi_0 \, \di  \varphi \sim M_{\rm Pl} \, H
\, h \, \di \varphi \; , \ee where $\phi_0(t)$ is the background
field configuration and $\varphi$ is the fluctuation, and we
assumed that the expansion of the universe is dominated by $\phi$.
The mixing term is suppressed by $H$, and at momenta and
frequencies much larger than $H$ it can be safely neglected. In
the case of a standard fluid one can repeat the same estimate, for
instance by means of the effective Lagrangian presented in
Ref.~\cite{fluids}. The result is exactly the same: the mixing between
sound waves and gravity is proportional to $H$. As a consequence,
the Jeans-instability time for a fluid that drives the expansion
of the universe is always of order $H^{-1}$. In our case the
situation is very different: we already stressed that all
quadratic terms we add to Eq.~(\ref{tadpoles_2}) gives
contributions to the matter stress-energy tensor that are linear
in the fluctuations, weighted by coefficients that are unrelated
to the background energy-density. This means that the mixing they
induce is completely unrelated to $H$, and can lead to
modifications of the fluctuation dispersion law at frequencies
parametrically larger than $H$.

In particular, the higher-derivative terms we are adding to cure
the gradient instability enhances the mixing of $\pi$ with gravity. The
larger we make $\Mbar^2$ in order to cure the gradient
instability, the bigger the effect of this mixing, and---if the
mixing tends to destabilize the system---the faster the
gravity-induced instability. We will see in the next section that
indeed when the full gravitational dynamics is taken into account
the system gets destabilized. 
In fact for frequencies much faster than $H$ and $\omega_{\rm grad}$
our system just reduces to the ghost condensate in Minkowski space, which features a Jeans-like instability \cite{Arkani-Hamed:2003uy}. In this regime we can estimate the rate of the gravity-induced instability in our system to be the same as that of the ghost condensate,
\be
\omega_{\rm Jeans} \sim \frac{\Mbar M^2}{M_{\rm Pl}^2}  \; , 
\ee
which is faster for larger $\Mbar^2$. In fact we get $\omega_{\rm grad} \, \omega_{\rm Jeans} \sim \dot H $. In this case it is obvious
that the best we can hope for is a compromise: the slower we make
the gradient instability, the faster the Jeans instability
becomes, and vice versa. As we discussed in section~\ref{example}
we want both kinds of instability to be slower than Hubble.
Therefore the parameters $M$ and $\Mbar$ must be chosen so that
the combination $\Mbar M^2 / M_{\rm Pl}^2$ lies in the interval
\be
\frac{\dot H}{H} \lesssim \frac{\Mbar M^2}{M_{\rm Pl}^2} \lesssim H \; .
\label{stabilitycondtion}
\ee This is really an interval only if
$\dot H$ is parametrically smaller than $H^2$. For $\dot H \sim
H^2$ the choice is highly constrained: $\Mbar M^2 /M_{\rm Pl}^2 \sim H$. Finally, for
$\dot H$ much larger than $H^2$ one of the two instabilities is
unavoidable.\\

So far we neglected a possible term
\be
\Delta S_{\rm matter} = 
- \int \! d^4 x \: \sqrt{- g} \: \hat M^3 \delta E^i {}_i \delta N \;.
\ee
Given the trasformation laws of $\delta N$ and
$\delta E_{ij}$, in terms of the $\pi$ field this operator corresponds to a three-derivative term of the form
$- \hat M^3 \dot \pi \frac{1}{a^2} \nabla^2 \pi$, apart from a negligible correction to the $\dot \pi^2$ term of
Eq.~(\ref{pi_dot}).
However upon integration by parts such a term can be rewritten as
\be
- \hat M^3 \dot \pi \frac{1}{a^2} \nabla^2 \pi \to  \hat M^3 \frac{1}{2 a^2} \frac{d}{dt}(\di_i \pi)^2
 \to - H \hat M^3 \frac{1}{a^2} (\di_i \pi)^2 \; ,
\ee
where again we are assuming that the time variation rate of the coefficient
$\hat M$ is generically of order $H$.
We therefore get a gradient energy term which,
although it is suppressed by $H$, can in principle compete
with that coming from the tadpoles, Eq.~(\ref{S_pi_tadpole}),
itself suppressed by $\dot H$.
In particular by choosing $\hat M^3$ larger than roughly
$\dot H M_{\rm Pl}^2 / H$ we can make the gradient energy positive.
But again we have to worry about the effect of mixing with gravity.
In analogy with the above discussion
such a mixing is unrelated with the background stress-energy
tensor and thus with $H$, and gets
enhanced when we take $\hat M^3$ larger and larger.
This makes the Jeans-like instability---if present---faster
and faster.
However there is a qualitative difference with respect to the previous
case: here there are no $k^4$ terms, so that the mixing with gravity
is relevant not only in the IR but at all scales. It is easy to check that the mixing terms become as important as the diagonal ones for $\hat M^3 \gtrsim H M_{\rm Pl}^2$. As we argued,
the gradient instability is cured for $\hat M^3 \gtrsim \dot H M_{\rm Pl}^2/H$. Then for $\dot H \ll H^2$
one can choose $\hat M^3/M_{\rm Pl}^2$ to lie in the same parametric range as before, Eq.~(\ref{stabilitycondtion}):
in such a range mixing with gravity is negligible and {\em the system is stable at all wavelengths},
with no need of higher-derivative terms in the UV and of the cosmological horizon in the IR.
As we will see in the next section, systematically taking into account the full gravitational dynamics shows that this parametric range for $\hat M^3$ is indeed the only one for which the system is stable.


In conclusion---as long as we restrict ourselves to systems that spontaneously break time translation
invariance---we verify the generality of the tension found in section~\ref{example}
between a violation
of the null energy condition and the stability of the system.  The system can be made completely stable only when $\dot{H}\ll H^2$. We are now going to confirm the results we got by a full calculation including gravitational effects.

\subsection{Full gravitational analysis \label{Paolo}}
In this section we want to write an explicit second order Lagrangian for scalar perturbations around an FRW background with $\dot H>0$ including the mixing of the scalar with gravity. This will show both the instabilities we discussed and the constraints on the model parameters to achieve stability.

Notice that the ADM variables $N$ and $N^i$ are not dynamical, 
i.e. the action does not contain their time derivative. They
should be thought as Lagrange multipliers: their equations of
motion are respectively the Hamiltonian and momentum constraints
of General Relativity. This means that we can solve for $N$ and
$N^i$ from their equations of motion and plug the result back into
the action, to get the Lagrangian for the scalar mode we are
interested in. The Lagrangian turns out to be more useful to study
instabilities than directly looking at the linearized equations of
motion. A ghost instability, where a system has the ``wrong'' sign
of the energy, does not show up if we study the linearized
equation of motions, but it will be manifest at the Lagrangian
level.

We choose unitary gauge, $\pi \equiv 0$. We have still to fix space diffeomorphisms and a convenient gauge
is given by \be \label{eq:gauge} h_{ij} = a(t)^2 \left[(1+2 \zeta)
\delta_{ij} +\gamma_{ij} \right]\;, \quad \partial_i
\gamma_{ij}\;, \quad \gamma_{ii} = 0 \;. \ee The matrix $\gamma$
describes tensor modes and in the following it will be neglected
as we are only interested in the dynamics of the scalar
perturbation $\zeta$. In this gauge the scalar perturbation is
given by an isotropic perturbations of the 3D metric at constant
$t$.

Let us now proceed to write an explicit Lagrangian for $\zeta$ at
second order. Postponing the discussion about the $\delta N \delta E^i {}_i$ term, we start from the action we constructed in the previous section,
\begin{eqnarray}
\label{eq:ADMaction} S & = & \int d^3x dt \, \sqrt{-g}
\left[\frac{M_{\rm Pl}^2}{2} \left(R^{(3)} + N^{-2} (E_{ij}
E^{ij}-E^i {}_i {}^2)\right)  - M_{\rm Pl}^2 \left(\frac{1}{N^2} \dot H(t) + 3 H(t)^2 +
 \dot H(t) \right) \right. \nonumber \\ & & \left. +\frac{M^4}{2}
\delta N^2 -\frac{\Mbar^2}{2}\delta E^i {}_i {}^2  \right] \;.
\end{eqnarray}
In the first line the first two terms reconstruct the full 4D
Ricci scalar, as in Eq.~(\ref{EH}). The other
terms of the first line are the tadpole terms as in Eq.~(\ref{tadpoles_2}), while on the second line we have contributions which are quadratic in the perturbations. For simplicity we neglect quadratic terms of the form $\delta E_{ij} \delta E^{ij}$, as we explicitly checked that they do not change the results, and we take $M$ and $\Mbar$ time independent.

The variation of the action with respect to $N$ gives the
Hamiltonian constraint \be \frac{M_{\rm Pl}^2}2 \left( R^{(3)} -N^{-2}
(E_{ij} E^{ij}-E^i {}_i {}^2) \right) -M_{\rm Pl}^2 \left(- N^{-2} \dot H +3 H^2 +
\dot H \right) + M^4 \delta N =0 \;, \ee while the variation with
respect to $N^i$ gives the momentum constraint 
\be M_{\rm Pl}^2 \hat \nabla_j
\left( N^{-1} (E^j {}_i - \delta^j_i E^k {}_k {} ) \right) - \Mbar^2
\hat \nabla_i E^k {}_k {} = 0 \;. 
\ee 
We are interested in the Lagrangian at
second order. Thus we have to solve these equations only at first
order in the perturbation $\zeta$, as second order terms would multiply,
once substituted back into the action, the unperturbed constraint
equations $\partial L/ \partial N$ and $\partial L/ \partial N^i$,
which vanish. Expressing the perturbations as $N = 1 + N_1$ and
$N^i = \partial_i \psi$ we can solve the two constraint equations
in terms of $\zeta$

\begin{eqnarray}
N_1 & = & \frac{4 M_{\rm Pl}^4 H \cdot \dot\zeta + 2 M_{\rm Pl}^2 \Mbar^2 \cdot \nabla^2 \zeta /a^2}{4 M_{\rm Pl}^4 H^2 + \Mbar^2 M^4} \\
\nabla^2\psi & = & \frac{(-18 M_{\rm Pl}^2 \Mbar^2 H^2 + 2 M_{\rm Pl}^2
M^4 - 4 M_{\rm Pl}^4 \dot H) \,\dot\zeta -4 M_{\rm Pl}^4 H \cdot
\nabla^2\zeta/a^2}{4 M_{\rm Pl}^4 H^2 + \Mbar^2 M^4} \;,
\end{eqnarray}
where we assumed $M, \Mbar \ll M_{\rm Pl}$.

The quadratic action for $\zeta$ is then obtained by substituting
these solutions into the original action Eq.~(\ref{eq:ADMaction}). After some integration by parts, the
result is given by \be\label{zetaaction} S = \int
d^3x \, dt  \, a^3(t) \left[ A(t) \,\dot\zeta^2 + B(t) \big(\frac{\partial_i}{a}
\zeta\big)^2 + C(t) \big(\frac{\partial^2}{a^2} \zeta\big)^2\right] \;.
\ee We have three operators quadratic in $\zeta$ with coefficients
which are time dependent and given by 
\be \label{eq:A} A(t) =
\frac{2\,M_{\rm Pl}^4\,\left(M^4 - 9 \Mbar^2\,H^2 - 2\,M_{\rm Pl}^2\,
\dot H \right) }
  {4\,M_{\rm Pl}^4\,H^2 + \Mbar^2  M^4 }
\ee
\begin{eqnarray}
B(t) =\frac{M_{\rm Pl}^2}{(4 M_{\rm Pl}^4 H^2 + \Mbar^2 M^4)^2} 
\left [-24 M_{\rm Pl}^6 \Mbar^2 H^4 \right. \!\!\! & + & \!\!\! \Mbar^2 (M^4 -2 M_{\rm Pl}^2 \dot H) (M^4 \Mbar^2 - 4 M_{\rm Pl}^4 \dot H) \nonumber \\  + \; 4 H^2 (M^4
M_{\rm Pl}^4 \Mbar^2 + 4 M_{\rm Pl}^8 \dot H) \!\!\! & - & \!\!\! \left. 8 M_{\rm Pl}^6 \Mbar^2 H
\ddot H \right] 
\end{eqnarray}
\be C(t) = - \frac{2 M_{\rm Pl}^4 \Mbar^2}{4 M_{\rm Pl}^4 H^2 +
\Mbar^2 M^4}\;. \ee Using the ADM approach we managed to get a
Lagrangian for the single relevant degree of freedom $\zeta$. It
contains all the information about the dynamics of the system, in
particular all the possible instabilities we are interested in.
Notice that we have no mass term for $\zeta$, but only terms
containing derivatives. The reason is quite general and it extends
beyond the quadratic order \cite{M}. From the definition Eq.~(\ref{eq:gauge})
it should be clear that if we can neglect spatial gradients of
$\zeta$, i.e. for a sufficiently long wavelength, this
variable cannot evolve in time, as it is equivalent to a constant
isotropic rescaling of the spatial coordinates. A mass term would
not allow this and it is therefore forbidden. Notice that the action for $\zeta$ allows, disregarding gradient terms, to study the homogeneous perturbations of the background 
and in particular to check if a background solution is an attractor.
From the time kinetic term of $\zeta$ in its action
Eq.~(\ref{zetaaction}) we immediately get the  solution 
\be \label{zetadot}
\dot\zeta = \frac{\rm const}{a(t)^3 A(t)} \;.
\ee
In this way, it is easy to check that slow-roll inflation and the ghost-condensate are attractors.

Let us now proceed to explicitly recover from this Lagrangian all
the qualitative results about instabilities of the previous
sections. As discussed, we are interested in time scales much
smaller than the Hubble time. This simplifies the algebra; for example in deriving the equations of motion
from the action above we can neglect the time variation
of the coefficients $A$, $B$ and $C$. An additional simplification
comes from the fact that for the validity of the effective field
theory description we must have $H \ll M$ and $H \ll \bar
M$. We thus reduce to a rather simple equation of motion 
\be
\ddot\zeta = \frac{(\Mbar^2 M^4 +4 M_{\rm Pl}^4 \dot H)k^2 - 2
M_{\rm Pl}^2 \Mbar^2 k^4}{2M_{\rm Pl}^2 (M^4-2 M_{\rm Pl}^2 \dot H)} \zeta \; ,
\ee 
where $k$ is the physical momentum.
As we discussed in the previous sections, in
Eq.~(\ref{eq:A}) we see that to avoid ghost-like instabilities we
have to take $M^4 > 2 M_{\rm Pl}^2 \dot H$. We can focus on the regime
$M^4 \gg 2 M_{\rm Pl}^2 \dot H$, because if the two terms are comparable
the instabilities are clearly worse. The resulting dispersion
relation captures all the qualitative features we discussed\footnote{In the limit we are discussing one can check that the inclusion of the $\delta E_{ij} \delta E^{ij}$ term just amounts to a redefinition of the constant $\Mbar$.} \be
\omega^2 = \frac{-(\Mbar^2 M^4 +4 M_{\rm Pl}^4 \dot H)k^2 + 2
M_{\rm Pl}^2 \Mbar^2 k^4}{2M_{\rm Pl}^2 M^4} \;. \ee 
The two terms proportional to $k^2$ describe the two instabilities we have discussed: the first, which is independent of $\dot H$, corresponds to the Jeans instability already present in Minkowski space, while the second leads to the gradient instability, weighted by $\dot H$. It is straightforward to find the unstable momentum $k$ with the fastest (imaginary) frequency; this defines the instability rate. We get
\be \label{favorite1}
\omega_{\rm max} = \left(\frac14 \frac{\Mbar M^2}{M_{\rm Pl}^2} + \dot H \frac{M_{\rm Pl}^2}{\Mbar M^2}\right)  \; .
\ee 
This formula unambiguously displays the complementarity of the two physically distinct instabilities we discussed; neither for $ \Mbar M^2/ M_{\rm Pl}^2 \to 0$ nor for $\Mbar M^2/ M_{\rm Pl}^2\to \infty$
do we end up with a stable system. 
One can try to minimize $\omega_{\rm max}$ by a proper choice of $\Mbar M^2/M_{\rm Pl}^2$. The rate of instability is minimized for
\be \label{favorite2}
\Mbar M^2/M_{\rm Pl}^2 = 2 \dot H^{1/2} \quad \rightarrow \quad \omega_{\rm max} = \dot H^{1/2} \;.
\ee
The full analysis therefore confirms what obtained in the previous sections. The instabilities can be pushed out of the horizon by a proper choice of the model parameters only if 
\be
\dot H \lesssim H^2 \;.
\ee

Let us now analyze the effect of the $\delta N \, \delta E^i {}_i {}$ operator. For simplicity we set to zero the $\delta E^i {}_i {}^2$ term, and consider the following action
\begin{eqnarray}
\label{eq:ADMactiontilde} S & = & \int d^3x dt \, \sqrt{-g}
\left[\frac{M_{\rm Pl}^2}{2} \left(R^{(3)} + N^{-2} (E_{ij}
E^{ij}-E^i {}_i {}^2)\right)  - M_{\rm Pl}^2 \left(\frac{1}{N^2} \dot H(t) + 3 H(t)^2 +
 \dot H(t) \right) \right. \nonumber \\ & & \left. +\frac{M^4}{2}
\delta N^2 - \hat M^3 \delta N \delta E^i {}_i {}  \right] \;.
\end{eqnarray}
The calculation follows exactly the same lines as before and we end up with an action for $\zeta$ of the form
\be S = \int
d^3x \, dt  \, a^3(t) \left[ \hat A(t) \,\dot\zeta^2 + \hat B(t) \big(\frac{\partial_i}{a}
\zeta\big)^2 \right] \;.
\ee
where the  the time-dependent coefficients are
\be \label{eq:tildeA} \hat A(t) =
\frac{M_{\rm Pl}^2 (- 12 M_{\rm Pl}^2 H \hat M^3 + 2 M^4 M_{\rm Pl}^2 - 4 M_{\rm Pl}^4 \dot H)}
  {(- \hat M^3 +2 M_{\rm Pl}^2 H)^2}
\ee
  \be \label{eq:tildeB} \hat B(t) =
\frac{M_{\rm Pl}^2 (\hat M^6 - 2 M_{\rm Pl}^2 H \hat M^3 + 4 M_{\rm Pl}^4 \dot H )}
  {(- \hat M^3 +2 M_{\rm Pl}^2 H)^2} \;.
\ee
Notice that, as expected, we have no $k^4$ terms, which in the previous case were coming from the operator $\delta E^i {}_i {}^2$. 
In the limit of time scales much smaller than $H^{-1}$ we get the dispersion relation 
\be
\omega^2 = \frac{-\hat M^6+2 M_{\rm Pl}^2 H \hat M^3 - 4 M_{\rm Pl}^4 \dot H}{2M_{\rm Pl}^2 M^4}k^2 \;. \ee 
Clearly the system is stable only if the coefficient of the $k^2$ term is positive. We can obtain this with a proper choice of $\hat M$, only if 
\be
\dot H \leq \frac{H^2}{4} \;.
\ee
In analogy with the previous case, there is a parametric window for $\hat M$ only if $\dot H \ll H^2$
\be
\frac{\dot H}{H} \lesssim \frac{\hat M^3}{M_{\rm Pl}^2} \lesssim H \;.
\ee

In conclusion, the  explicit calculation confirms what we argued in the previous section. Both with the $\delta E^i {}_i {}\delta E^i {}_i {}$ and with the $\delta N \delta E^i {}_i {}$ terms we can have stability only if $\dot H \lesssim H^2$. In a generic effective action both quadratic terms will be present at the same time but at this point it is clear that this cannot change the qualitative picture.



\section{\label{applications} Applications}
In this section we give some illustrative examples of
non-standard cosmological histories that are now allowed as the NEC can be violated without introducing pathologies.


\subsection{Today's Acceleration\label{acc}}
After the surprising discovery that the universe is presently accelerating
(in the sense that $\ddot a >0$), it is natural to consider the even more
exotic possibility that the Hubble rate is growing with time
($\dot H = \ddot a/a -(\dot a/a)^2 >0$).
It has become common to parameterize the present expansion in terms of the
equation of state   parameter $w \equiv p/\rho$, where $p$ and $\rho$ are
respectively the present pressure and energy density of the universe.
The violation of the null energy condition is equivalent in this language
to the inequality $w < -1$.
Present data require $w \gtrsim-1.2$ \cite{Spergel:2006hy,Seljak:2006bg}, leaving room for observing $w<-1$.

The simple toy model in section \ref{example} gives an example of a model
with $w < -1$ with no fine tuning or instability.
More general models can easily be written using the
formalism of section \ref{effective}.
As already discussed, to avoid instabilities the model parameters must
satisfy the inequalities
 \be\label{condnow}
-(1+w) H \lesssim \frac{\Mbar M^2}{M_{\rm Pl}^2} \lesssim H \; ,
\ee 
or analogous ones in the presence of a
$\hat M^3 \, \delta N \delta E^i {}_i {}$ term.
If we demand that instabilities are parametrically suppressed,
we must have
$\dot{H} \ll H^2$, and therefore $|w + 1| \ll 1$.
It is phenomenologically more interesting to consider the case where
$w$ is near the lower experimental limit $|w + 1| \sim 0.2$.
In this case, the instability is not parametrically suppressed, and we
expect that the new degree of freedom has an unstable mode.
We stress again that this is a long-wavelength instability like the
familiar Jeans instability for matter, and not a breakdown of the
effective theory.
The presence of unstable modes in dark energy in models with $|w + 1| \sim 1$ is very general,
and gives a possible new observable handle on these models.
We leave the details for future work.

We want to stress that this way of explaining a possible observation of
$w < -1$ relies on a genuine violation of the null energy condition:
the {\em Einstein frame} metric has $\dot H >0$.
This is completely different from other ``phenomenological" approaches
that mimic a violation of the null energy condition.
For example one way to get $w < -1$ is to couple matter to a conformally
rescaled metric $f(\phi) g_{\mu\nu}$, where $g_{\mu\nu}$ is the metric
in Einstein frame and $\phi$ is an evolving scalar field.
Experiments will be sensitive to the way this new metric evolves,
so that an observation of $w < -1$ would not imply a violation of
the null energy condition. Notice however that this kind of models is severely constrained by fifth force experiments, see
for example  Refs.~\cite{Carroll:1998zi,fluids}.
Other approaches \cite{Lue:2004za,Das:2005yj} mimic the presence of a
dark energy component with $w<-1$ while keeping $\dot H <0$:
the total equation of state of dark matter and dark energy has $w> -1$. 


\subsection{Starting the Universe \label{starting}}
If the NEC is satisfied, a very general property of an
expanding universe is that it always evolves from a state with high energy
density towards a state with a lower one.
This implies that any effective field theory description will eventually break
down if we go sufficiently backwards in the past.
Now that we have shown that it is possible to violate the NEC
without introducing patologies, we can consider the possibility
that the universe ``starts" from a very low energy state,
gains energy in the expansion,
and eventually reaches an high energy state from which the standard
cosmological evolution begins.
In particular, the universe could
{\em approach flatness in the asymptotic past}, $H\to 0$ for $t\to -\infty$.

Notice that in any such model, independently of the details of the evolution, the present causal horizon is always infinite,
since the universe expanded for an infinite proper time.
This gives a potential alternative approach
to standard inflation for solving the homogeneity problem of the universe.

It is straightforward to implement this in the framework of the general
effective action presented in section~\ref{effective},
specifying for example in Eq.~(\ref{theeffectiveaction}) a suitable
function $H(t)$.
For instance, if we take the scaling solution
$H(t)\simeq \alpha /|t|$ for $t\to-\infty$
with $\alpha \gg 1$, we approach flatness in the past,
while keeping the required parametric separation 
\be
\dot{H}\simeq \frac\alpha {|t|^2} \ll \frac{\alpha^2}{ |t|^2}\simeq H^2\, .
\ee
The instabilities are under control if 
\be \label{startuniversecondition}
\frac{\dot{H}}{H}\lesssim \frac{\Mbar(t)M(t)^2}{M^2_{\rm Pl}}\lesssim H\ .
\ee
Since $H(t)\to 0$ for $t\to -\infty$, we see that the model parameters must
change with time in order to keep these conditions satisfied.
As we discussed, since the system is not time-translationally invariant,
it is generic that the coefficients of the effective Lagrangian
$M,\tilde{M},\hat M,\dots$ explicitly depend on time with a time scale of
order $H$.
There is therefore no difficulty in making the system stable at all times.

It is easy to give an explicit realization of this scenario
in a scalar field model similar to that of section \ref{example}.
Besides an appropriate choice of the potential, an explicit $\phi$
dependence must also be present in the derivative terms.
At the level of the $\pi$ action, this translates into a time dependence
of the parameters, so that with an appropriate choice,
the condition Eq.~(\ref{startuniversecondition}) can be satisfied.
Notice that the explicit $\phi$ dependence is compatible with the fact that the
shift symmetry on $\phi$ is already broken by the potential.
To give an example, we start from a Lagrangian of the form
Eq.~(\ref{eq:Lghost}), which admits the solution $\phi = t$.
We choose the minimum of $P$ to be at zero, e.g.
\be \label{L1}
P(X) = \frac 18 \, (X - 1)^2 \; ,
\ee 
so that in the absence of a potential term the solution for the metric
is Minkowski space.
Since we want $H$ to vary slowly with time,
the stress-energy tensor must be dominated by the potential.
To reproduce $H = \alpha/|t|$ we need
\be \label{L2}
V \sim \frac{\alpha^2 M_{\rm Pl}^2 }{\phi^2}\;.
\ee 
As in section~\ref{example}, this will slightly displace
$\dot \phi$ from the minimum of $P$, but 
this perturbation is small at early times and the corresponding correction
to the stress-energy tensor is negligible.
This solution is approaching a ``big rip'' singularity as $t \to 0$,
so at some negative $t$ it will break down.
In particular, at $t \sim - \alpha^{1/2} M_{\rm Pl}/M^2$ we would have
$\dot \pi_0 \sim 1$, and the unperturbed solution cannot be trusted any more.
This  corresponds to a maximum Hubble rate of order
$H_{\rm max} \sim \alpha^{1/2} M^2/ M_{\rm Pl}$.
Before this happens, we can match the solution to a standard FRW phase.

To complete the construction of the model, we must
introduce higher derivative terms.
To satisfy the condition Eq.~(\ref{startuniversecondition}) at all times,
we need $\Mbar$ to depend on time, which is accomplished by making the
4-derivative terms depend on $\phi$ as \footnote{In this way we introduce a large hierarchy between $M$ and $\Mbar$ which may be difficult to justify at the level of effective field
theory. Alternatively both scales could move together remaining in the allowed range Eq.~(\ref{startuniversecondition}). In this toy model, we stick to the simplest case in which only 
$\Mbar$ is time dependent.}
\be \label{L3}
\Delta {\cal L} =
\Mbar ^2 _0  \, \frac{\phi_0^2}{\phi^2} \, (\Box \phi)^2 \; .
\ee
One can check that the contribution of this term to the stress-energy tensor is negligible
at sufficiently early times, up to Hubble rates of order $M^2/M_{\rm Pl}$.
The $\phi$-dependence of the potential and of the $(\Box \phi)^2$, besides making the mass scales time-dependent, introduces
small changes in the dynamics of the perturbations.
For example, masses and additional $(\di \pi)^2$ terms will be generated,
but they will be negligible.
This can be simply understood from the fact that for large $\alpha$ the rise of $H$ is very slow and the system approaches the ghost condensate in de Sitter space.
%

In summary, putting together the Lagrangian terms Eqs.~(\ref{L1}--\ref{L3}) we have an  explicit model for a universe which starts expanding from a zero-curvature state without developing any instability. 
The point is not to present a particular compelling cosmological model,
but to stress that it is possible to build models where the present universe
evolved from a state which is asymptotically flat in the past.

\subsection{The Eternally Expanding Cyclic Universe \label{periodicexample}}

Given the two scenarios above, one is tempted to put them together to build an 
eternally expanding universe with a cyclic evolution.
We can imagine, as in section \ref{acc}, that the present expansion has $\dot H >0$. This means that in the future the universe will be inflating with a larger and larger energy density.
At a certain point this energy can be converted to matter and radiation as in a conventional inflation model to start a new FRW-like evolution
\cite{Arkani-Hamed:2003uz,Senatore:2004rj}.
Then one is lead to connect the present acceleration with the inflationary
phase which occurred in our past, to build a periodic universe that goes
through this cosmological history many times.

An illustrative example is based on the model of section \ref{example}.
At the present epoch, the field rolls up the potential giving a
super-accelerating phase with growing energy.
The reheating to a new FRW phase can be achieved by a sudden drop of
the potential.
If we imagine that the field $\phi$ is periodic
(see Fig.~\ref{fig:periodicthepotential}), we obtain a periodic
cosmological evolution in which the present acceleration is the beginning
of a super-inflationary phase identical to the one that is responsible
for the structure formation in our universe.

As already discussed in section \ref{example}, the linear potential is
technically natural because the shift symmetry is unbroken in the absence
of gravity.
The sharp drop of the potential responsible for reheating breaks the shift
symmetry even in the absence of gravity, and will therefore induce more general
terms that break the shift symmetry through radiative corrections.
However, the breaking of the shift symmetry is localized in $\phi$,
so it is technically natural for the potential to be linear a distance
$\Delta \phi \gsim M$ away from the sharp drops.
%

The model can be made stable for the entire cosmological history.
If at the present epoch Eq.~(\ref{condnow}) is satisfied, we know that
we are safe from instabilities while the scalar rolls up the potential.
Looking backwards in time from the present epoch, we enter into a matter
dominated and then radiation dominated phase
(see Fig.~\ref{fig:periodicthepotential}).
As $H$ is larger then nowadays, the Jeans instability, whose rate is
independent of $H$, is under control.
Also the gradient instability is not an issue since its rate is
proportional to $V'/H$ and it thus becomes smaller and smaller in the past.

 
%


\begin{figure}[th!!]
\begin{center}
\includegraphics[width=11cm]{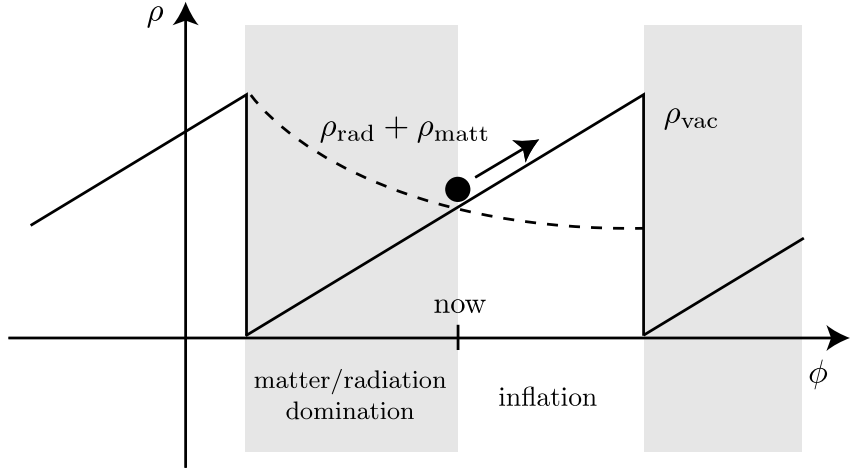}
\caption{\label{fig:periodicthepotential} \small Schematic
representation of the periodic model presented in section \ref{periodicexample}.}
\end{center}
\end{figure}

If we take this model seriously, 
it gives a striking relationship between the present equation of
state $w$ and the tilt of the inflationary spectrum $n_s$:
they both depend on the constant slope of the potential $V'$.


Following Ref.~\cite{Arkani-Hamed:2003uz,Senatore:2004rj}
the spectrum of density perturbations goes as
\begin{equation}
P_\zeta \sim\left(\frac{H_I}{M}\right)^{\frac{5}{2}}\label{pr}
\end{equation}
where $H_I$ is the Hubble rate during the inflationary
phase. The slow variation of $H_I$ gives a tilt:  
\begin{equation}
n_s-1\simeq \frac{5}{12} \,\frac{V'}{M^2_{\rm Pl} H^3_I} \;.
\end{equation}
On the other hand, the current acceleration of the universe
would be characterized by an equation of state $w$ equal to
\begin{equation}
1+w=-\frac{V'}{9 M^2_{\rm Pl}H^3_0} \; ,
\end{equation}
where $H_0$ is the current Hubble rate. We thus get the relation
\begin{equation}
n_s-1=- \frac{15}{4} \, \left(\frac{H_0}{H_I}\right)^3(1+w) \; .
\end{equation}
Given the constraints on $w$, the tilt of the spectrum is predicted to be extremely small and practically impossible to measure.
Moreover, to be able to test this relationship one should determine $H_I$.
The production of gravitational waves does not help, since the energy scale
$M$ is constrained to be too low.
In principle $M$ could be determined from gravitational experiments or
through the direct coupling of this scalar sector to the standard model,
so that $H_I$ could be fixed through Eq.~(\ref{pr}).
If the detection of a negative tilt by the WMAP experiment
\cite{Spergel:2006hy} is confirmed this specific model will be ruled
out.

If we allow the scale $M$ to be a function of time we do not have
constraints coming from the present day modification of gravity. In
this case the amplitude of gravitational waves could be sufficiently
large to be observed. Notice that these gravitational waves would have
a {\em blue} spectrum, a striking signature of violation of the NEC. 

In conclusion we have shown how the violation of the null energy condition opens up the possibility of relating the present and past evolution of our universe and of building periodic cosmological histories.

\subsection{The Big Bounce\label{bounce}}
The null energy condition implies that the cosmological evolution
in the absence of spatial curvature cannot bounce from a contracting
to an expanding phase, since we need $\dot H>0$ to make $H$ change sign.
We expect that bouncing cosmologies can be realized in our scenarios.
However, as $H=0$ at the bounce, one could wonder whether the instabilities
can be kept under control.
Obviously the real issue is whether the instabilities have sufficient
time to grow.
We will see that it is possible to control instabilities by making the
bounce sufficiently fast. 
   

Let us take the following simple evolution of the scale factor around the bounce (see figure \ref{fig:bounce})
\begin{equation}
a(t)=1+\frac{t^2}{2T^2} \;,
\end{equation}
so that for $t \ll T$ 
\begin{equation}
H(t)\simeq\frac{t}{T^2} \label{Hbounce} \;.
\end{equation}
\begin{figure}[t!]
\begin{center}
\includegraphics[width=8cm]{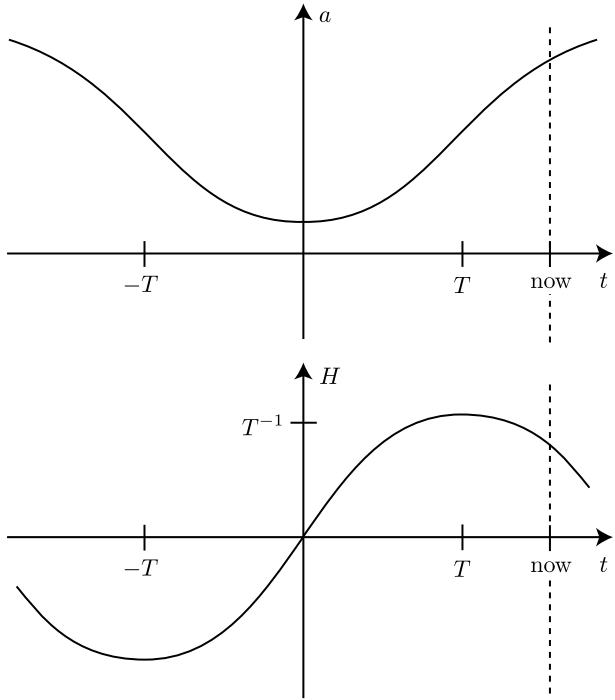}
\caption{\label{fig:bounce} \small Schematic dependence of the scale factor $a(t)$ and of the Hubble rate $H(t)$ in the bouncing model of section \ref{bounce}, assuming
that the bounce is matched to standard $\dot H <0$ phases both for $t< -T$ and for $t>T$.}
\end{center}
\end{figure}
A system that realizes this bounce is described for example by the
$\pi$ action Eq.~(\ref{theeffectiveaction}), with $H(t)$ as above.
The stability condition Eq.~(\ref{stabilitycondtion}) shows the potential problem.
The Hubble rate, which is crucial in stabilizing the Jeans and gradient
instabilities, goes to zero at the bounce. 

Let us see how the problem can be solved.
For simplicity we assume that the parameters $M$ and $\Mbar$ are time-independent. 
Notice that during the bounce $\dot H$ is constant, $\dot H \simeq 1/T^2$.
As we have shown in section~\ref{effective},
for a given $\dot H$ the
instability rate is minimized for $\Mbar M^2/M_{\rm Pl}^2 = 2 \dot H^{1/2}$,
which gives
\be
\omega_{\rm inst} = \dot H^{1/2} = \frac1T \; .
\ee
This choice makes the rate of gradient and Jeans instability the same; see
Eqs.~(\ref{favorite1}--\ref{favorite2}).

In this case the only time scale during the bounce is $T$.
In particular at $t \sim \pm T$ the Hubble rate is  $H(\pm T) \sim \pm 1/T$.
Therefore the period for which the instability is faster than $H$ only
lasts a time of order $T$, which is one instability time.
That is, the instability does not have time to develop. There is, however, another form of instability one has to worry about during the contracting phase.
In a contracting universe anisotropic contributions to the Friedmann
equation blue-shift like $a^{-6}$.
This means that in order for a contracting solution not to be destroyed
by anisotropic metric perturbations the contraction must satisfy
$\dot{H}<- 3 H^2$ \cite{Erickson:2003zm}.
Of course when the bounce phase sets in at $t \sim -T$ this condition is
violated, but since the scale factor only varies by a factor of order unity
during the bounce, the anisotropies will only grow by an order one factor,
which obviously is not a problem.

Outside the $|t|<T$ window, one can match the bounce to a contracting phase for $t<-T$ and to an expanding one for $t>T$. The expanding phase can be a standard decelerated FRW evolution, 
similarly to what happens in ekpyrotic/cyclic scenarios, or a super-accelerating phase with $\dot H >0$, which can make $H$ as large as we like.
For negative $t$ one is forced to consider a $\dot H < 0$ phase to avoid growing anisotropic perturbations.  Notice that as $H$ decrease going to negative $t$ (and also for positve $t$
if $\dot H < 0$) the Jeans instability is active. This can be cured as we did in section~\ref{starting},
e.g.~by postulating that $\Mbar^2$ has a suitable time-dependence that makes
the Jeans instability rate slower than $H$.
This is certainly consistent at the level of the effective field theory for the fluctuations we developed in the previous section.
In the language of the scalar $\phi$, this would correspond to a 
particular $\phi$ dependence of the higher derivative part of the Lagrangian.
Finding the explicit form of such a Lagrangian 
is out of the scope of the present paper.

We conclude that it is possible to realize a FRW solution in
which the universe bounces from a contracting phase to an
expanding one, without developing catastrophic instabilities.
To our knowledge, this is the first case in which
a bouncing solution is realized in a situation where the
importance of all the higher derivative terms is controlled in an
effective field theory expansion at low energies.

In this controllable setup it is easy to follow the evolution of cosmological perturbations from the contracting to the expanding phase. This is crucial for the experimental viability of 
ekpyrotic/cyclic scenarios \cite{justin}. Since our bounce satisfies the general hypotheses of Ref.~\cite{matias}, it unfortunately leads to a  non scale-invariant prediction for density perturbations.

\section{Conclusions and Outlook\label{outlook}}
In this paper we have shown that at the level of effective field theory there
is no obstruction in constructing systems that violate the
null energy condition (NEC).
In particular, the instabilities that plague a very broad class of
NEC-violating systems \cite{fluids} can be completely avoided.
We use this to construct cosmological models with
$\dot{H} > 0$ that do not have any instabilities.

These results depend crucially on the interplay between two
potential types of instabilities.
The first are gradient instabilities that occur very generally
whenever the NEC is violated.
These can be avoided due to higher-derivative terms in the
effective theory.
However, increasing the higher-derivative terms increases the
effect of another potential instability, the Jeans instability
that arises from the attractive nature of gravity.
The Jeans instability can be avoided if the time scale for
the instability is longer than the Hubble time.
Demanding that both of these instabilities are absent
requires
\be
\label{conditionconclusion}
\dot{H} \lsim H^2 \; .
\ee
This means that we cannot get an arbitrarily large violation
of the NEC.
These results are derived both in a specific model
(a deformation of ghost condensation \cite{Arkani-Hamed:2003uy} deformed by
the addition of a potential \cite{Arkani-Hamed:2003uz,Senatore:2004rj})
and in a very general effective field theory analysis of
adiabatic scalar fluctuations about an arbitrary FRW
background.

We emphasize that if Eq.~(\ref{conditionconclusion}) is violated
the effective field theory does not break down, but it contains long wavelength exponentially 
growing modes.
These might be interesting for cosmology:
for example, if the present acceleration of the universe has
$w = p /\rho < -1$, then it becomes possible that they might have shown up around our present epoch.

We constructed a number of explicit examples to show how this can
lead to interesting alternative cosmological scenarios.
We constructed a simple explicit model that gives $w < -1$ today.
We also presented a model in which the universe starts from
Minkowski space in the distant past, giving a possible alternative
to standard inflation for solving the horizon problem.
Putting these together, we consider an eternally expanding
cyclic model in which the present accelerating phase is the beginning
of inflation in the next cycle.
This opens up the intriguing possibility that measurements today
can give information about the inflationary phase that gave rise
to the structure we observe today.
Finally, we constructed a model which has a smooth ``bounce''
connecting a contracting phase to the present expanding phase.
Although bouncing cosmologies have been previously considered in
the literature, we believe ours is the first example in which
the bounce is under theoretical control.
These illustrate some of the phenomenologically interesting
possibilities opened up by violation of the NEC.
We leave detailed analysis of these ideas to future work.

We close with some theoretical questions. 
It would be interesting to understand whether the requirement $\dot H \ll H^2$ to have a parametric suppression of instabilities
is completely generic or not. Models involving more degrees of freedom could be completely stable even when the NEC is strongly violated.

While the results of this paper are under control in a systematic
effective field theory expansion, it is natural to ask whether this
particular effective field theory can be embedded in a UV complete
theory such as renormalizable quantum field theory or string theory.
In particular, the somewhat exotic features of the ghost condensate
(e.g. Lorentz breaking and complicated nonlinear dynamics
\cite{Arkani-Hamed:2003uy}) may make one suspicious that the ghost condensate
lies in the ``swampland'' of effective field theories without UV
completions \cite{Vafa:2005ui, Arkani-Hamed:2006dz}.
In fact, in Ref.~\cite{adams} it was shown that a large class of 
consistent effective field theories (including some interesting
modifications of gravity) have no Lorentz invariant UV completion,
essentially because they allow configurations in which signals
travel faster than light \cite{adams}.
The present model does not suffer from this problem, since the
excitations of the ghost condensate have a dispersion relation
$\omega^2 \simeq k^4/M^2$, and therefore travel much
\emph{slower} than light.
However, it is true that there is currently no known UV completion
of the model;
for an interesting recent attempt, see Ref.~\cite{ian}.

Despite these open questions about the full consistency of the model
at all energy scales, we believe that
the phenomenological possibilities opened up
by violation of the NEC are well worth further exploration.

\section*{Acknowledgments}
We would like to thank Nima Arkani-Hamed, Sergei Dubovsky, Riccardo Rattazzi, and
Matias Zaldarriaga for useful discussions.
M.A.L.~and A.N.~would like to thank the Galileo Galilei Institute for Theoretical Physics for
hospitality and INFN for partial support during the completion of this work. A.N.~would also like to thank
the Abdus Salam International Centre for Theoretical Physics for hospitality and support.
M.A.L.~is supported by the National Science Foundation
grant PHY-0354401 and the University of Maryland Center for
Particle and String Theory.
L.S.~is supported in part by funds provided by the
U.S.~Department of Energy (D.O.E.) under
cooperative research agreement DF-FC02-94ER40818.

\section*{Note added}
In a recent paper \cite{rubakov} a NEC-violating model was presented. It involves a vector field as well as a scalar. Like ours, the model has no UV---i.e., arbitrarily fast---instabilities, but only exponential instabilities below a critical momentum $k_c$. However, not even in the `slow roll' regime $\dot H \ll H^2$ can the instability rate $\omega_c$ be made slower than the Hubble rate.
What prevents the system from being disrupted is then that fluctuations exit the horizon, so that instabilities only have a finite amount of time to develop before being damped by Hubble friction.
On the other hand, in our case when $\dot H \ll H^2$ the system can be made completely stable.


\footnotesize
\parskip 0pt

\begin{thebibliography}{nn}

\bibitem{Spergel:2006hy}
  D.~N.~Spergel {\it et al.},
  ``Wilkinson Microwave Anisotropy Probe (WMAP) three year results:
  Implications for cosmology,''
  astro-ph/0603449.

\bibitem{Seljak:2006bg}
  U.~Seljak, A.~Slosar and P.~McDonald,
  ``Cosmological parameters from combining the Lyman-alpha forest with CMB,
  galaxy clustering and SN constraints,''
  astro-ph/0604335.

\bibitem{riccardo}
R.~Rattazzi, ``A new dimension at ultra large scales and its price'', talk at SUSY2K, unpublished, http://wwwth.cern.ch/susy2k/susy2kfinalprog.html .
  
\bibitem{ghostconstraints2}
  J.~M.~Cline, S.~Jeon and G.~D.~Moore,
  ``The phantom menaced: Constraints on low-energy effective ghosts,''
  Phys.\ Rev.\ D {\bf 70}, 043543 (2004)
  [hep-ph/0311312].

\bibitem{Holdom:2004yx}
  B.~Holdom,
  ``Accelerated expansion and the Goldstone ghost,''
  JHEP {\bf 0407}, 063 (2004)
  [hep-th/0404109].

\bibitem{fluids}
  S.~Dubovsky, T.~Gregoire, A.~Nicolis and R.~Rattazzi,
  ``Null energy condition and superluminal propagation,''
  JHEP {\bf 0603}, 025 (2006)
  [hep-th/0512260].

\bibitem{Arkani-Hamed:2003uy}
  N.~Arkani-Hamed, H.~C.~Cheng, M.~A.~Luty and S.~Mukohyama,
  ``Ghost condensation and a consistent infrared modification of gravity,''
  JHEP {\bf 0405}, 074 (2004)
  [hep-th/0312099].
  
\bibitem{sergio}
  S.~L.~Dubovsky,
  ``Phases of massive gravity,''
  JHEP {\bf 0410}, 076 (2004)
  [hep-th/0409124].

\bibitem{justin}
  J.~Khoury,
  ``A briefing on the ekpyrotic / cyclic universe,''
  astro-ph/0401579 and references therein.
  
\bibitem{Senatore:2004rj}
  L.~Senatore,
  ``Tilted ghost inflation,''
  Phys.\ Rev.\ D {\bf 71}, 043512 (2005)
  [astro-ph/0406187].

\bibitem{Arkani-Hamed:2005gu}
  N.~Arkani-Hamed, H.~C.~Cheng, M.~A.~Luty, S.~Mukohyama and T.~Wiseman,
  ``Dynamics of gravity in a Higgs phase,''
  hep-ph/0507120.

\bibitem{weinberg}
 S.~Weinberg,
 ``Adiabatic modes in cosmology,''
  Phys.\ Rev.\ D {\bf 67}, 123504 (2003)
  [astro-ph/0302326].

\bibitem{Arkani-Hamed:2002sp}
  N.~Arkani-Hamed, H.~Georgi and M.~D.~Schwartz,
  ``Effective field theory for massive gravitons and gravity in theory space,''
  Annals Phys.\  {\bf 305}, 96 (2003)
  [hep-th/0210184].

\bibitem{Arnowitt:1962hi}
  R.~Arnowitt, S.~Deser and C.~W.~Misner,
  ``The Dynamics Of General Relativity,''
  gr-qc/0405109.

\bibitem{Hsu:2004vr}
  S.~D.~H.~Hsu, A.~Jenkins and M.~B.~Wise,
  ``Gradient instability for $w<-1$,''
  Phys.\ Lett.\ B {\bf 597}, 270 (2004)
  [astro-ph/0406043].

\bibitem{M}
  J.~M.~Maldacena,
  ``Non-Gaussian features of primordial fluctuations in single field
  inflationary models,''
  JHEP {\bf 0305}, 013 (2003)
  [astro-ph/0210603].

\bibitem{Carroll:1998zi}
  S.~M.~Carroll,
  ``Quintessence and the rest of the world,''
  Phys.\ Rev.\ Lett.\  {\bf 81}, 3067 (1998)
  [astro-ph/9806099].

   
\bibitem{Lue:2004za}
  A.~Lue and G.~D.~Starkman,
  ``How a brane cosmological constant can trick us into thinking that $w <
  -1$,''
  Phys.\ Rev.\ D {\bf 70}, 101501 (2004)
  [astro-ph/0408246].
  
\bibitem{Das:2005yj}
  S.~Das, P.~S.~Corasaniti and J.~Khoury,
  ``Super-acceleration as signature of dark sector interaction,''
  Phys.\ Rev.\ D {\bf 73}, 083509 (2006)
  [astro-ph/0510628].

\bibitem{Arkani-Hamed:2003uz}
  N.~Arkani-Hamed, P.~Creminelli, S.~Mukohyama and M.~Zaldarriaga,
  ``Ghost inflation,''
  JCAP {\bf 0404} (2004) 001
  [hep-th/0312100].
  
\bibitem{Erickson:2003zm}
 See for example  J.~K.~Erickson, D.~H.~Wesley, P.~J.~Steinhardt and N.~Turok,
  ``Kasner and mixmaster behavior in universes with equation of state $w \ge
  1$,''
  Phys.\ Rev.\ D {\bf 69}, 063514 (2004)
  [hep-th/0312009] and reference therein.

 \bibitem{matias}
  P.~Creminelli, A.~Nicolis and M.~Zaldarriaga,
  ``Perturbations in bouncing cosmologies: Dynamical attractor vs scale
  invariance,''
  Phys.\ Rev.\ D {\bf 71}, 063505 (2005)
  [hep-th/0411270].

\bibitem{Vafa:2005ui}
  C.~Vafa,
  ``The string landscape and the swampland,''
  hep-th/0509212.

\bibitem{Arkani-Hamed:2006dz}
  N.~Arkani-Hamed, L.~Motl, A.~Nicolis and C.~Vafa,
  ``The string landscape, black holes and gravity as the weakest force,''
  hep-th/0601001.

\bibitem{adams}
  A.~Adams, N.~Arkani-Hamed, S.~Dubovsky, A.~Nicolis and R.~Rattazzi,
  ``Causality, analyticity and an IR obstruction to UV completion,''
  hep-th/0602178.

\bibitem{ian}
  M.~L.~Graesser, I.~Low and M.~B.~Wise,
  ``Towards a high energy theory for the Higgs phase of gravity,''
  Phys.\ Rev.\ D {\bf 72}, 115016 (2005)
  [hep-th/0509180].

\bibitem{rubakov}
  V.~A.~Rubakov,
  ``Phantom without UV pathology,''
  hep-th/0604153.




\end{thebibliography}
\end{document}